\documentclass[a4paper,final,onecolumn]{article}%
\usepackage{amsfonts}
\usepackage{amsmath}
\usepackage{amssymb}
\usepackage{graphicx}
\usepackage{version}
\usepackage{accents}%
\providecommand{\U}[1]{\protect\rule{.1in}{.1in}}

\begin{document}

\title{Quantization in the neighborhood of the initial instant}
\author{{\small Z. Belhadi\thanks{Email : zahir.belhadi@univ-bejaia.dz}}\\{\small Laboratoire de physique th\'{e}orique, Facult\'{e} des sciences
exactes,}\\{\small Universit\'{e} de Bejaia, 06000 Bejaia, Alg\'{e}rie.}}
\date{}
\maketitle

\subsection*{Abstract}

\qquad In this paper, we present an approach to quantize singular systems.
This is an extension of the constant integration method \cite{nous} which is
applicable only for the case of exactly solvable systems. In our approach, we
determine Dirac brackets at the initial instant with the help of Taylor
expansion, and using their covariance, we deduce the brackets at any time. We
also illustrate our approach by studying two systems: the self-dual model and
the Dirac spinorial field.

\textbf{Keywords} : quantization, singular systems, integration constants,
Liouville operator, Taylor expansion. \vspace{-0.6cm}

\bigskip

\section{Introduction}

\qquad Quantization is a transition procedure from the classical deterministic
formulation to the probabilistic quantum description of physical phenomena.
The question of quantization is still of actuality, as it has not received so
far any definitive nor satisfactory answer, especially for theories like
general relativity which resist any attempt at a quantum description.

One procedure to deal with this question is the canonical quantization based
on the knowledge of fundamental variables brackets, which are necessary to
deduce the commutators of the corresponding quantum operators. In the regular
case where all phase space variables are independent, we speak of Poisson
brackets well known in analytical mechanics, otherwise, Dirac brackets
generalize them to the case of singular Lagrangians \cite{Dirac,erer}
characterezed by the presence of constraints. To calculate these brackets, the
Dirac approach consists first, on the determination and classification of all
the constraints, then inverting the matrix of these constraints to obtain the
desired brackets.

The method of Faddeev-Jackiw \cite{FJ} on the other hand, is based on the
linearization of the Lagrangian with respect to the velocities, to get first
order equations of motion. After the determination of all constraints and the
inversion of the symplectic matrix, one accesses to the brackets necessary for
canonical quantization. \

Sometimes, improvements can be made by using for instance the analytical
solution of the equations of motion to obtain the different brackets
\cite{nous}. This approach can be straightforward and reproducible, but needs
a complete general solution with all the independent integration constants.

The aim purpose of this paper is to generalize this last approach, initially
developed for exactly solvable systems, and adapt it to the case of
non-integrable systems, using Taylor polynomial expansion as an approximation
to the solution in the neighborhood of the initial conditions. Here, the role
played by the integration constants will be attributed to the initial
conditions, and the Dirac brackets will be deduced directly from the brackets
computed at the initial instant, without introducing the concept of
constraints and their classification.

We will illustrate the correctness of our approach by studying two models of
the field theory : the self-dual model and the spinorial field. In both cases,
the Dirac brackets needed for the canonical quantization are calculated by
just approximating the solution of the equations of motion with a first order
Taylor expansion.

\section{Method of integration constants (CI)}

\qquad Consider a classical system described by a singular autonomous
Lagrangian $L(q,\dot{q})$ where $q=(q_{1},...,q_{N})$ are the generalized
coordinates and $\dot{q}=(\dot{q}_{1},...,\dot{q}_{N})$ the generalized
velocities. Suppose we have the (general) analytical solutions $q(t)=q(t,C)$
of the Euler-Lagrange equations and the momenta$\ p(t)=p(t,C)$ $\left(
p_{i}=\frac{\partial L}{\partial\dot{x}}\right)  $, where $C=(C_{1}%
,C_{2},...,C_{M})$ is the set of integration constants. For
constrained\textbf{\footnote{$M=2N$ corresponds to an unconstrained system.
For the constrained systems, each second class constraint eliminates one
variable.}} systems we have $M<2N.$

From the analytical solutions of the equations of motion we can write the
Hamiltonian\footnote{one can also obtain the Hamiltonian by putting the
solutions into the Legendre transformation $H=\sum_{i}\frac{d\tilde{q}%
_{i}(t,C)}{dt}\tilde{p}_{i}(t,C)-L\left(  \tilde{q}_{i}(t,C),\frac{d\tilde
{q}_{i}(t,C)}{dt}\right)  .$ In this way, one do not have to inverse the
momenta with respect to the velocities.} as
$H(q(t),p(t))=H(q(t,C),p(t,C))=\tilde{H}(C).$ In the CI method \cite{nous}, we
determine the integration constants brackets with the help of the following
property%
\begin{equation}
\left\{
\begin{array}
[c]{c}%
\frac{\partial}{\partial t}q_{i}(t,C)=%
{\textstyle\sum\limits_{k,l=1}^{M}}
\{C_{k},C_{l}\}\frac{\partial q_{i}}{\partial C_{k}}\frac{\partial H}{\partial
C_{l}}\ \ \ \ i=1...N\vspace{0.2cm}\\
\frac{\partial}{\partial t}p_{i}(t,C)=%
{\textstyle\sum\limits_{k,l=1}^{M}}
\{C_{k},C_{l}\}\frac{\partial p_{i}}{\partial C_{k}}\frac{\partial H}{\partial
C_{l}}\ \ \ \ i=1...N.
\end{array}
\right.  \label{hyhyhy}%
\end{equation}
These $2N$ equations contain $M(M-1)/2$ unknown brackets $\{C_{k},C_{l}\},$
with $k,l=1...M$. Our method consists in the determination of the brackets
$\{C_{k},C_{l}\}$ via simple identifications$.$ In the case where the general
solution depends on arbitrary functions due to gauge symmetry, we must fix the
gauge before computing any bracket. Using the brackets $\{C_{k},C_{l}\},$ we
can deduce the brackets $\left\{  q_{i},q_{j}\right\}  $, $\left\{
p_{i},p_{j}\right\}  $ and $\left\{  q_{i},p_{j}\right\}  $ more easily than
with any other existing approach. In fact, when we replace the fundamental
variables\textbf{ }$q_{i}$\textbf{ }and \textbf{ }$p_{i}$\textbf{ }by the
solution $q_{i}(t,C)$\textbf{ }and\textbf{ }$p_{i}(t,C),$\textbf{ }we have the
property%
\begin{equation}
\left\{  f,g\right\}  =%
{\textstyle\sum\limits_{k,l=1}^{M}}
\{C_{k},C_{l}\}\frac{\partial f}{\partial C_{k}}\frac{\partial g}{\partial
C_{l}}.
\end{equation}
The weakness of this method is the need to know the analytical solution with
all the independent constants, which is not always possible, especially in the
presence of interaction terms. In the next section, we will solve this problem
and adapt this approach to non-integrable systems.

\section{Proof of Dirac brackets covariance in time}

\qquad It is known in quantum mechanics that the commutators in the Heisenberg
picture taken at equal times have the same form as the commutators in the
Schr\"{o}dinger picture calculated at the initial instant. More precisely, the
equal-time commutation relations conserve their form
\begin{equation}
\left[  A,B\right]  =C\text{ \ \ }\Rightarrow\text{ \ \ }\left[
A(t),B(t)\right]  =C(t).
\end{equation}
In this section, we will give the classical equivalent of this property. In
other words, knowing Dirac brackets (generalizing the Poisson ones) of the
fundamental variables $\xi(t)=(q(t),p(t))$ at the initial instant, it is
possible to deduce them at any time
\begin{equation}
\{\tilde{\xi}_{I},\tilde{\xi}_{J}\}=\Theta_{IJ}(\tilde{\xi})\text{
\ \ \ }\Rightarrow\text{ \ \ \ }\{\xi_{I}(t),\xi_{J}(t)\}=\Theta_{IJ}(\xi(t)),
\label{zerzer}%
\end{equation}
where the $\tilde{\xi}_{I}$ are the initial conditions and $\{\tilde{\xi}%
_{I},\tilde{\xi}_{J}\}$ their brackets at the initial instant\footnote{Even in
analytical mechanics, when one speaks of the bracket $\left\{  x,p\right\}
=1,$ one means the equal time bracket $\left\{  x(t),p(t)\right\}  =1$
associated to the instant $t.$ So, it is logical to talk about the initial
time bracket $\left\{  x(0),p(0)\right\}  =\left.  \left\{  x(t),p(t)\right\}
\right\vert _{t=0}=1.$ Indeed, all the brackets propreties are valid only at
equal times.}.

Let's start by the evolution equation of a fundamental variable determined by
the Hamilton equation
\begin{equation}
\frac{d\xi_{K}}{dt}=\{\xi_{K},H\}=\{\xi_{K},\xi_{J}\}\frac{\partial
H}{\partial\xi_{J}}=\{\xi_{I},\xi_{J}\}\frac{\partial H}{\partial\xi_{J}}%
\frac{\partial}{\partial\xi_{I}}\xi_{K}.
\end{equation}
With the help of the Liouville operator $G=G(\xi_{I},\xi_{J})$ defined by%
\begin{equation}
G(\xi_{I},\xi_{J})=\{\xi_{I},\xi_{J}\}\frac{\partial H}{\partial\xi_{J}}%
\frac{\partial}{\partial\xi_{I}}%
\end{equation}
we can deduce all the derivatives of $\xi_{K}(t)$ as follows%
\begin{equation}
\frac{d\xi_{K}}{dt}=G\xi_{K}\ \ \ \ \Rightarrow\ \ \ \ \frac{d^{n}\xi_{K}%
}{dt^{n}}=G^{n}\xi_{K}.
\end{equation}
Using Taylor series expansion near the initial conditions $\tilde{\xi}_{K}%
=\xi_{K}(0),$ one obtains%
\begin{align}
\xi_{K}(t)  &  =\sum_{n}\left.  \frac{d^{n}\xi_{K}}{dt^{n}}\right\vert
_{t=0}\frac{t^{n}}{n!}=\sum_{n}\left.  G^{n}\xi_{K}\right\vert _{t=0}%
\frac{t^{n}}{n!}\nonumber\\
\xi_{K}(t)  &  =\left(  \sum_{n}\tilde{G}^{n}\frac{t^{n}}{n!}\right)
\tilde{\xi}_{K}, \label{sebbac}%
\end{align}
where
\begin{equation}
\tilde{G}=\left.  G\right\vert _{t=0}=\left.  \{\xi_{I},\xi_{J}\}\right\vert
_{t=0}\left.  \frac{\partial H}{\partial\xi_{J}}\right\vert _{t=0}\left.
\frac{\partial}{\partial\xi_{I}}\right\vert _{t=0}=\{\tilde{\xi}_{I}%
,\tilde{\xi}_{J}\}\frac{\partial\tilde{H}}{\partial\tilde{\xi}_{J}}%
\frac{\partial}{\partial\tilde{\xi}_{I}}%
\end{equation}
is the Liouville operator associated to the initial conditions $\tilde{\xi},$
and $\tilde{H}=H(\tilde{\xi})$ is the Hamiltonian which is conserved because
the Lagrangian is autonomous$.$We can deduce the formal solution
(\ref{sebbac})
\begin{equation}
\xi_{K}(t)=e^{t\tilde{G}}\tilde{\xi}_{K}\text{ \ \ \ where \ \ \ \ }%
e^{t\tilde{G}}=\sum_{n}\tilde{G}^{n}\frac{t^{n}}{n!}. \label{drfdrf}%
\end{equation}
The same reasoning allows us to show that any analytical function $f(\xi)$ can
be written as%
\begin{equation}
f(\xi)=e^{t\tilde{G}}f(\tilde{\xi}), \label{uki}%
\end{equation}
where
\begin{equation}
\tilde{G}f(\tilde{\xi})=\{\tilde{\xi}_{I},\tilde{\xi}_{J}\}\frac
{\partial\tilde{H}}{\partial\tilde{\xi}_{J}}\frac{\partial}{\partial\tilde
{\xi}_{I}}f(\tilde{\xi})=\{f(\tilde{\xi}),\tilde{H}\}.
\end{equation}
From the previous relation and using the identity of Jacobi $\left\{  \left\{
\tilde{\xi}_{I},\tilde{\xi}_{J}\right\}  ,\tilde{H}\right\}  +\left\{
\left\{  \tilde{\xi}_{J},\tilde{H}\right\}  ,\tilde{\xi}_{I}\right\}
+\left\{  \left\{  \tilde{H},\tilde{\xi}_{I}\right\}  ,\tilde{\xi}%
_{J}\right\}  =0$, we obtain the Leibniz rule
\begin{equation}
\tilde{G}\{\tilde{\xi}_{I},\tilde{\xi}_{J}\}=\{\tilde{G}\tilde{\xi}_{I}%
,\tilde{\xi}_{J}\}+\{\tilde{\xi}_{I},\tilde{G}\tilde{\xi}_{J}\}.
\end{equation}
The successive application of $\tilde{G}$ on the bracket $\{\tilde{\xi}%
_{I},\tilde{\xi}_{J}\}$ leads to the well-known general Leibniz rule
\begin{equation}
\tilde{G}^{k}\{\tilde{\xi}_{I},\tilde{\xi}_{J}\}=\sum_{s=0}^{k}\frac
{k!}{s!\left(  k-s\right)  !}\{\tilde{G}^{s}\tilde{\xi}_{I},\tilde{G}%
^{k-s}\tilde{\xi}_{J}\}. \label{seresere}%
\end{equation}
Our goal is to use the formal solution (\ref{drfdrf}) to demonstrate the
proposition (\ref{zerzer}). Indeed,
\begin{align}
\{\xi_{I}(t),\xi_{J}(t)\}  &  =\{e^{t\tilde{G}}\tilde{\xi}_{I},e^{t\tilde{G}%
}\tilde{\xi}_{J}\}\nonumber\\
&  =\sum_{(n,m)\in%
\mathbb{N}
^{2}}\frac{t^{n}}{n!}\frac{t^{m}}{m!}\{\tilde{G}^{n}\tilde{\xi}_{I},\tilde
{G}^{m}\tilde{\xi}_{J}\}. \label{ghghgh}%
\end{align}
The elements of $%
\mathbb{N}
^{2}$ are
\[%
\begin{tabular}
[c]{cccccccc}%
$(0,0)\vspace{0.2cm}$ & $\hspace{-0.3cm}(0,1)$ & $\hspace{-0.3cm}(0,2)$ &
$\hspace{-0.3cm}(0,3)$ & $\hspace{-0.3cm}(0,4)$ & $\hspace{-0.3cm}(0,5)$ &
$\hspace{-0.3cm}(0,6)$ & $\hspace{-0.3cm}\cdots$\\
$(1,0)\vspace{0.2cm}$ & $\hspace{-0.3cm}(1,1)$ & $\hspace{-0.3cm}(1,2)$ &
$\hspace{-0.3cm}(1,3)$ & $\hspace{-0.3cm}(1,4)$ & $\hspace{-0.3cm}(1,5)$ &
$\hspace{-0.3cm}(1,6)$ & $\hspace{-0.3cm}\cdots$\\
$(2,0)\vspace{0.2cm}$ & $\hspace{-0.3cm}(2,1)$ & $\hspace{-0.3cm}(2,2)$ &
$\hspace{-0.3cm}(2,3)$ & $\hspace{-0.3cm}(2,4)$ & $\hspace{-0.3cm}(2,5)$ &
$\hspace{-0.3cm}(2,6)$ & $\hspace{-0.3cm}\cdots$\\
$(3,0)\vspace{0.2cm}$ & $\hspace{-0.3cm}(3,1)$ & $\hspace{-0.3cm}(3,2)$ &
$\hspace{-0.3cm}(3,3)$ & $\hspace{-0.3cm}(3,4)$ & $\hspace{-0.3cm}(3,5)$ &
$\hspace{-0.3cm}(3,6)$ & $\hspace{-0.3cm}\cdots$\\
$(4,0)\vspace{0.2cm}$ & $\hspace{-0.3cm}(4,1)$ & $\hspace{-0.3cm}(4,2)$ &
$\hspace{-0.3cm}(4,3)$ & $\hspace{-0.3cm}(4,4)$ & $\hspace{-0.3cm}(4,5)$ &
$\hspace{-0.3cm}(4,6)$ & $\hspace{-0.3cm}\cdots$\\
$(5,0)\vspace{0.2cm}$ & $\hspace{-0.3cm}(5,1)$ & $\hspace{-0.3cm}(5,2)$ &
$\hspace{-0.3cm}(5,3)$ & $\hspace{-0.3cm}(5,4)$ & $\hspace{-0.3cm}(5,5)$ &
$\hspace{-0.3cm}(5,6)$ & $\hspace{-0.3cm}\cdots$\\
$(6,0)$ & $\hspace{-0.3cm}(6,1)$ & $\hspace{-0.3cm}(6,2)$ & $\hspace
{-0.3cm}(6,3)$ & $\hspace{-0.3cm}(6,4)$ & $\hspace{-0.3cm}(6,5)$ &
$\hspace{-0.3cm}(6,6)$ & $\hspace{-0.3cm}\cdots$\\
$\vdots$ & $\hspace{-0.3cm}\vdots$ & $\hspace{-0.3cm}\vdots$ & $\hspace
{-0.3cm}\vdots$ & $\hspace{-0.3cm}\vdots$ & $\hspace{-0.3cm}\vdots$ &
$\hspace{-0.3cm}\vdots$ & $\hspace{-0.3cm}\ddots$%
\end{tabular}
\ \ \ \ \
\]
Following the antidiagonal, one can see that $%
\mathbb{N}
^{2}=\cup_{k=0}^{\infty}S_{k}$ where $S_{k}=\{(s,k-s)\in%
\mathbb{N}
^{2}\backslash$ $0\leq s\leq k\}=\{(0,k),(1,k-1),...,(k,0)\}.$ If $C(n,m)$ is
a coefficient depending on $n$ and $m,$ then
\begin{align}
\sum_{{\small (n,m)\in%
\mathbb{N}
}^{2}}C(n,m)  &  =\sum_{k=0}^{\infty}C(0,k)+...+C(k,0)\\
&  =\sum_{k=0}^{\infty}\sum_{s=0}^{k}C(s,k-s)
\end{align}
We use this property to write (\ref{ghghgh}) in the form
\begin{equation}
\{\xi_{I}(t),\xi_{J}(t)\}=\sum_{k=0}^{\infty}\sum_{s=0}^{k}\frac{t^{k}%
}{s!\left(  k-s\right)  !}\{\tilde{G}^{s}\tilde{\xi}_{I},\tilde{G}^{k-s}%
\tilde{\xi}_{J}\}
\end{equation}
Finally, we achieve our demonstration with the help of (\ref{seresere}) and
(\ref{uki})%
\begin{equation}
\{\xi_{I}(t),\xi_{J}(t)\}=\sum_{k=0}^{\infty}\frac{t^{k}}{k!}\tilde{G}%
^{k}\Theta_{IJ}(\tilde{\xi})=\Theta_{IJ}(\xi(t))
\end{equation}

We have thus shown that, in the autonomous case, the knowledge of brackets at
the initial instant is sufficient to deduce them at any later instant without
even talking about the constraints. This result is of crucial importance
because it will allow us to use the CI method to determine the desired
brackets between the initial conditions by solving the equations of motion in
infinitesimal time interval around the initial instant, and subsequently
obtain the equal time brackets using the covariance of Dirac brackets.

As a first example, consider the Lagrangian
\begin{equation}
L=\frac{\dot{x}^{2}}{2}+\left(  z+e^{-x}\right)  \dot{y}-z\frac{x^{2}}%
{2}\text{ }\Rightarrow\text{ }H=\frac{p_{x}^{2}}{2}+\frac{x^{2}z}{2}%
\end{equation}
The Euler-Lagrange equations lead to the equations of motion
\begin{equation}
\left\{
\begin{array}
[c]{l}%
\ddot{x}=x\dot{y}-zx\\
\dot{z}=-e^{-x}\text{ }\dot{x}\\
\dot{y}=\frac{x^{2}}{2}%
\end{array}
\right.  \ \ \text{ \ \ and \ }\ \ \left\{
\begin{array}
[c]{l}%
p_{x}=\dot{x}\\
p_{y}=z+e^{-x}\\
p_{z}=0
\end{array}
\right.  \label{EL}%
\end{equation}
The initial conditions associated to this system are $x(0)=X,$ $y(0)=Y,$
$z(0)=Z,$ $p_{x}(0)=P_{x},$ $p_{y}(0)=P_{y}$ and $p_{z}(0)=P_{z}.$ This system
of equations is not analytically solvable, so, we simply take the Taylor
expansion of their solution. In other words, let us write the unknown
functions as power series in the form $x_{i}(t)=x_{i}(0)+\sum_{n=1}^{\infty
}a_{i}(n)\frac{t^{n}}{n!},$ then insert them into the previous differential
system to determine the coefficients $a_{i}(n)$ in term of the initial
conditions. After a straightforward calculation, we get the relations \
\begin{equation}
\left\{
\begin{array}
[c]{l}%
x(t)=X+P_{x}\text{ }t+\left(  \frac{X^{3}}{2}-ZX\right)  \frac{t^{2}}%
{2}+O(t^{3})\\
z(t)=Z-e^{-X}P_{x}\text{ }t+O(t^{2})\\
y(t)=Y-\frac{X^{2}}{2}t+O(t^{2})
\end{array}
\right.
\end{equation}
and%
\begin{equation}
\left\{
\begin{array}
[c]{l}%
p_{x}=P_{x}+\left(  \frac{X^{3}}{2}-ZX\right)  t+O(t^{2})\\
p_{y}=\frac{X^{2}}{2}+Z+O(t)\\
p_{z}=0.
\end{array}
\right.
\end{equation}
It is clear that the only independent initial conditions are $(X,Y,Z,P_{x})$
while $P_{z}=0$ and $P_{y}=X+Z$. $\ $The Hamiltonian is a conserved quantity
because we are in the presence of an autonomous Lagrangian, so $H=H|_{t=0}%
=\frac{P_{x}^{2}}{2}+\frac{X^{2}Z}{2}.$

At this point, we use the Hamilton equations near the initial instant $\left.
\dot{\xi}_{i}\right\vert _{t=0}=\left\{  \left.  \xi_{i}\right\vert
_{t=0},H\right\}  $ for $\xi_{i}\in\{x,y,z,p_{x}\}$ to obtain the different
brackets at this instant. Thus,%
\begin{equation}
\dot{x}|_{t=0}=\left\{  x|_{t=0},H\right\}  \Rightarrow P_{x}=\{X,\frac
{P_{x}^{2}}{2}+\frac{X^{2}Z}{2}\}.
\end{equation}
The properties of the brackets allow us to deduce
\begin{equation}
P_{x}=\{X,P_{x}\}P_{x}+\{X,Z\}\frac{X^{2}}{2}.
\end{equation}
In the same way, we obtain the following relations
\begin{equation}
\left\{
\begin{tabular}
[c]{l}%
$-e^{-X}P_{x}=\{Z,P_{x}\}P_{x}+\{Z,X\}ZX$\\
$-\frac{X^{2}}{2}=\{Y,P_{x}\}P_{x}+\{Y,Z\}\frac{X^{2}}{2}+\{Y,X\}XZ$\\
$\frac{X^{2}}{2}-ZX=\{P_{x},Z\}\frac{X^{2}}{2}+\{P_{x},X\}XZ.$%
\end{tabular}
\ \ \right.
\end{equation}
After identification, we get the brackets among the initial conditions
\begin{align}
\{X,P_{x}\}  &  =1\ \ \ \ \{X,Z\}=0\ \ \ \ \ \{Z,P_{x}\}=-e^{-X}\\
\{Y,P_{x}\}  &  =0\ \ \ \ \ \{Y,Z\}=-1\ \ \ \ \ \ \ \{Y,X\}=0.
\end{align}
This result is valid at any time according to the above theorem, and thus%
\begin{align}
\{x,p_{x}\}  &  =1\ \ \ \ \{x,z\}=0\ \ \ \ \ \{z,p_{x}\}=-e^{-x}\\
\{y,p_{x}\}  &  =0\ \ \ \ \ \{y,z\}=-1\ \ \ \ \ \ \ \{x,y\}=0.
\end{align}
The other brackets can be easily obtained from the equations $p_{y}=z+e^{-x}$
and $p_{z}=0.$ To verify the validity of this result, it suffices to ensure
that the Hamilton equations obtained with these brackets are equivalent to the
Euler-Lagrange equations (\ref{EL}).

This example clearly shows that the Dirac brackets of a non-integrable
singular system can be deduced directly with the help of the Taylor polynomial
expansion without invoking the notion of constraints, and this constitutes the
major strength of this approach.

\section{SD model}

\qquad The self-dual (SD$)$ model, originally introduced by Townsend, Pilch
and Van Nieuwenhuizen as a three dimensional "square root" of Proca equation
is described by the Lagrangian density \cite{town}
\begin{equation}%
\mathcal{L}%
=-\frac{1}{2m}\varepsilon^{\mu\nu\rho}f_{\mu}\partial_{\nu}f_{\rho}+\frac
{1}{2}f_{{}}^{\mu}f_{\mu},
\end{equation}
where the positive real parameter $m$ can be seen as a mass. The signature of
our metric is $(+,-,-)$ and the Levi-Civita tensor $\varepsilon^{\mu\nu\rho}$
is equal to $1$ if $(\mu,\nu,\rho)$ is an even permutation of $(0,1,2)$, $-1$
if it is an odd permutation, and all other components are $0$. We define also
$\varepsilon^{ij}=\varepsilon^{0ij}$ where $i,j\in\{1,2\}.$

The purpose of this section is to apply our approach to this model. Indeed,
Euler-Lagrange equations lead to the equations of motion
\begin{equation}
\varepsilon^{\mu\nu\rho}\partial_{\nu}f_{\rho}-mf_{{}}^{\mu}%
=0\ \ \ \Leftrightarrow\ \ \ \left\{
\begin{array}
[c]{c}%
\dot{f}^{1}=-mf^{2}-\partial_{1}f^{0}\\
\dot{f}^{2}=mf^{1}-\partial_{2}f^{0}\\
-\partial_{1}f^{2}+\partial_{2}f^{1}=mf^{0}.
\end{array}
\right.  \label{df}%
\end{equation}
The momenta are $\pi^{0}=0$, $\pi^{1}=-\frac{1}{2m}f_{2}$ and $\pi^{2}%
=\frac{1}{2m}f_{1}$ and the Hamiltonian takes the form%
\begin{equation}
H=\int dx^{2}\left(  -\frac{1}{2}f_{{}}^{\mu}f_{\mu}+\frac{1}{m}f^{0}%
\partial^{1}f^{2}-\frac{1}{m}\text{ }f^{0}\partial^{2}f^{1}\right)  ,
\end{equation}
where we used the relation $\varepsilon_{ij}f_{i}\partial_{j}f_{0}%
=\partial_{j}\left(  \varepsilon_{ij}f_{i}f_{0}\right)  -\varepsilon_{ij}%
f_{0}\partial_{j}f_{i},$ and eliminated the 2-divergence term $\partial
_{j}\left(  \varepsilon_{ij}f_{i}f_{0}\right)  .$

\bigskip

Let's now start with the initial conditions $f^{\mu}(t=0,x^{1},x^{2})=F^{\mu
}(x^{1},x^{2})$ and $\pi^{\mu}(t=0,x^{1},x^{2})=\Pi^{\mu}(x^{1},x^{2}).$ From
the equation (\ref{df}), taken near the initial instant $t=0$,%

\begin{equation}
\ \left\{
\begin{array}
[c]{l}%
F^{0}=\frac{1}{m}\left(  \partial_{2}F^{1}-\partial_{1}F^{2}\right) \\
\dot{f}^{1}|_{t=0}=-mF^{2}-\partial_{1}F^{0}\\
\dot{f}^{2}|_{t=0}=mF^{1}-\partial_{2}F^{0}%
\end{array}
\right.  \text{ and }\left\{
\begin{array}
[c]{l}%
\Pi^{0}=0\\
\Pi^{1}=\frac{1}{2m}F^{2}\\
\Pi^{2}=-\frac{1}{2m}F^{1}.
\end{array}
\right.
\end{equation}
Only two initial conditions $F^{1}$ and $F^{2}$ are independent$.$ The
Hamiltonian is a constant ($H=H(0)$) and takes the form
\begin{align}
H  &  =\int dx^{2}{\Huge (}\frac{\left(  F^{1}\right)  ^{2}}{2}+\frac{\left(
F^{2}\right)  ^{2}}{2}\nonumber\\
&  +\frac{1}{2m^{2}}\left(  \left(  \partial_{2}F^{1}\right)  ^{2}+\left(
\partial_{1}F^{2}\right)  ^{2}-2\partial_{1}F^{2}\text{ }\partial_{2}%
F^{1}\right)  {\Huge ).}%
\end{align}

To determine the bracket $\left\{  F^{1}(\vec{x}),F^{2}(\vec{x}^{\prime
})\right\}  ,$ we impose the Hamilton equation near the initial instant
\begin{equation}
\dot{f}^{1}|_{t=0}=\left\{  f^{1}|_{t=0},H\right\}  ,
\end{equation}
where
\begin{equation}
\dot{f}^{1}|_{t=0}=-mF^{2}-\partial_{1}F^{0}=-mF^{2}-\frac{1}{m}\left(
\partial_{1}\partial_{2}F^{1}-\partial_{1}^{2}F^{2}\right)  , \label{8965}%
\end{equation}
and
\begin{align}
\left\{  F^{1},H\right\}   &  =\int dx^{\prime2}{\Large (}F^{\prime2}\left\{
F^{1},F^{\prime2}\right\} \nonumber\\
&  +\frac{1}{m^{2}}(\partial_{1}^{\prime}F^{^{\prime}2}\left\{  F^{1}%
,\partial_{1}^{\prime}F^{^{\prime}2}\right\}  -\partial_{1}^{\prime}%
F^{\prime2}{\Large \{}F^{1},\partial_{2}^{\prime}F^{\prime1}{\Large \}}%
\nonumber\\
&  {\Large -\{}F^{1},\partial_{1}^{\prime}F^{\prime2}\text{ }{\Large \}}%
\partial_{2}^{\prime}F^{\prime1}){\Large )}, \label{4588}%
\end{align}
where $F^{\prime\mu}=F^{\mu}(\vec{x}^{\prime})$ and $\partial_{i}^{\prime
}=\frac{\partial}{\partial x^{\prime i}}.$ But just comparing the equations
(\ref{8965}) and (\ref{4588}), we obtain the wanted bracket
\begin{equation}
\left\{  F^{1}(\vec{x}),F^{2}(\vec{x}^{\prime})\right\}  =-\frac{1}{m}%
\delta(\vec{x}-\vec{x}^{\prime}).
\end{equation}
Therefore $\left\{  F^{1},\partial_{1}^{\prime}F^{\prime2}\right\}  =-\frac
{1}{m}\partial_{1}^{\prime}\delta(\vec{x}-\vec{x}^{\prime})$ and
${\Large \{}F^{1},\partial_{2}^{\prime}F^{\prime1}{\Large \}}=\partial
_{2}^{\prime}{\Large \{}F^{1},F^{\prime1}{\Large \}}=0$. The time covariance
of the brackets allows us to deduce the bracket%
\begin{equation}
\left\{  f^{i}(\vec{x},t),f^{j}(\vec{x}^{\prime},t)\right\}  =-\frac
{\varepsilon^{ij}}{m}\delta(\vec{x}-\vec{x}^{\prime}).
\end{equation}
The other brackets can be obtained using to the relations $f^{0}=\frac{1}%
{m}\left(  \partial_{2}f^{1}-\partial_{1}f^{2}\right)  ,$ $\pi^{0}=0$,
$\pi^{1}=\frac{1}{2m}f^{2}$ et $\pi^{2}=-\frac{1}{2m}f^{1}.$

This result is similar to that found in literature \cite{jk,autre}, and this
shows the correctness of our method in the case of the self dual model.
\bigskip\

\section{Dirac free Field}

\qquad The Dirac field Lagrangian density is of the form $%
\mathcal{L}%
=i\overline{\Psi}\gamma^{\mu}\partial_{\mu}\Psi-m\overline{\Psi}\Psi,$ where
$\Psi=(\Psi_{1},\Psi_{2},\Psi_{3},\Psi_{4})^{T}$ are complex odd Grassman
fields and $\overline{\Psi}=(\Psi^{\ast})^{T\text{ }}\gamma^{0}.$ The Dirac
matrices ($\gamma^{0},\gamma^{i}$) are related to the usual matrices
($\beta,\alpha^{i}$) by the relations $\beta=$\ $\gamma^{0}$ and $\alpha
^{i}\ =\gamma^{0}\gamma^{i}=\beta\gamma^{i}.$ The Euler-Lagrange equations
$\partial_{\mu}\frac{\vec{\partial}%
\mathcal{L}%
}{\partial\left(  \partial_{\mu}\Psi\right)  }=\frac{\vec{\partial}%
\mathcal{L}%
}{\partial\Psi}$ and $\partial_{\mu}\frac{\vec{\partial}%
\mathcal{L}%
}{\partial\left(  \partial_{\mu}\overline{\Psi}\right)  }=\frac{\vec{\partial}%
\mathcal{L}%
}{\partial\overline{\Psi}}$ expressed in term of the left derivative
$\vec{\partial}$ lead to the well-known Dirac equations $i\partial_{t}%
\Psi=-i\alpha^{i}\partial_{i}\Psi+m\beta\Psi$ and $i\partial_{t}\overline
{\Psi}=-i\partial_{i}\overline{\Psi}$ $\alpha^{i}-m\overline{\Psi}\beta.$

Suppose that the initial conditions are $\Psi_{a}(\vec{x},0)=\Phi_{a}(\vec
{x})$ where $a\in\{1,2,3,4\}.$ Then, to satisfy the equation $i\partial
_{t}\Psi=-i\alpha^{i}\partial_{i}\Psi+m\beta\Psi$ near the initial instant
$t_{0}=0$, the first order Taylor series expansion of $\Psi$ must be of the
form%
\begin{equation}
\Psi_{a}(\vec{x},t)=\Phi_{a}(\vec{x})-\left(  \alpha_{ab}^{i}\partial_{i}%
\Phi_{b}(\vec{x})+im\beta_{ab}\Phi_{b}(\vec{x})\right)  t+O(t^{2}).
\end{equation}
The canonical Hamiltonian $H=\int d\vec{x}^{\prime}{\Large (}-i\Phi_{c}^{\ast
}(\vec{x}^{\prime})\alpha_{cb}^{i}\partial_{i}^{\prime}\Phi_{b}(\vec
{x}^{\prime})+m\Phi_{c}^{\ast}(\vec{x}^{\prime})\beta_{cb}\Phi_{b}(\vec
{x}^{\prime}){\Large )}$ is an even Grassmann conserved quantity. Let us now
compute $\left.  \dot{\Psi}_{a}(\vec{x},t)\right\vert _{t=0}$ and $\left\{
\left.  \Psi_{a}(\vec{x},t)\right\vert _{t=0},H\right\}  .$ We have
\begin{align}
\left.  \dot{\Psi}_{a}(\vec{x},t)\right\vert _{t=0}  &  =\int d\vec{x}%
^{\prime}{\Large (}-\delta_{ac}\alpha_{cb}^{i}\partial_{i}^{\prime}\Phi
_{b}(\vec{x}^{\prime})\delta(\vec{x}^{\prime}-\vec{x})\nonumber\\
&  \text{ \ \ \ }-im\delta_{ac}\beta_{cb}\Phi_{b}(\vec{x}^{\prime})\delta
(\vec{x}^{\prime}-\vec{x}){\Large ),}%
\end{align}
and%
\begin{align}
\left\{  \left.  \dot{\Psi}_{a}(\vec{x},t)\right\vert _{t=0},H\right\}   &
=\int d\vec{x}^{\prime}{\Large (}-i\alpha_{cb}^{i}\left\{  \Phi_{a}(\vec
{x}),\Phi_{c}^{\ast}(\vec{x}^{\prime})\partial_{i}^{\prime}\Phi_{b}(\vec
{x}^{\prime})\right\} \nonumber\\
&  \text{ \ \ \ \ }+m\beta_{cb}\left\{  \Phi_{a}(\vec{x}),\Phi_{c}^{\ast}%
(\vec{x}^{\prime})\Phi_{b}(\vec{x}^{\prime})\right\}  {\Large ).}%
\end{align}
Using the equation $\left.  \dot{\Psi}_{a}(\vec{x},t)\right\vert _{t=0}$
=$\left\{  \left.  \Psi_{a}(\vec{x},t)\right\vert _{t=0},H\right\}  $, and we
get, after identification, the relations%
\begin{align}
\left\{  \Phi_{a}(\vec{x}),\Phi_{c}^{\ast}(\vec{x}^{\prime})\Phi_{b}(\vec
{x}^{\prime})\right\}   &  =-i\delta_{ac}\Phi_{b}(\vec{x}^{\prime})\delta
(\vec{x}^{\prime}-\vec{x})\\
\left\{  \Phi_{a}(\vec{x}),\Phi_{c}^{\ast}(\vec{x}^{\prime})\partial
_{i}^{\prime}\Phi_{b}(\vec{x}^{\prime})\right\}   &  =-i\delta_{ac}%
\partial_{i}^{\prime}\Phi_{b}(\vec{x}^{\prime})\delta(\vec{x}^{\prime}-\vec
{x}).
\end{align}
The brackets of odd Gassmann (fermionic) variables ($O_{1},O_{2},O_{3})$
satisfy \cite{casa} the property $\left\{  O_{1},O_{2}O_{3}\right\}  =\left\{
O_{1},O_{2}\right\}  O_{3}-O_{2}\left\{  O_{1},O_{3}\right\}  ,$ and therefore%
\begin{align}
\delta_{ac}\Phi_{b}(\vec{x}^{\prime})\delta(\vec{x}^{\prime}-\vec{x})  &
=i\left\{  \Phi_{a}(\vec{x}),\Phi_{c}^{\ast}(\vec{x}^{\prime})\right\}
\Phi_{b}(\vec{x}^{\prime})\nonumber\\
&  -i\Phi_{c}^{\ast}(\vec{x}^{\prime})\left\{  \Phi_{a}(\vec{x}),\Phi_{b}%
(\vec{x}^{\prime})\right\}  ,
\end{align}
and%
\begin{align}
\delta_{ac}\partial_{i}^{\prime}\Phi_{b}(\vec{x}^{\prime})\delta(\vec
{x}^{\prime}-\vec{x})  &  =i\left\{  \Phi_{a}(\vec{x}),\Phi_{c}^{\ast}(\vec
{x}^{\prime})\right\}  \partial_{i}^{\prime}\Phi_{b}(\vec{x}^{\prime
})\nonumber\\
&  -i\Phi_{c}^{\ast}(\vec{x}^{\prime})\left\{  \Phi_{a}(\vec{x}),\partial
_{i}^{\prime}\Phi_{b}(\vec{x}^{\prime})\right\}  .
\end{align}
Finally, one obtains after identification, the relations
\begin{equation}
\left\{  \Phi_{a}(\vec{x}),\Phi_{c}^{\ast}(\vec{x}^{\prime})\right\}
=-i\delta_{ac}\delta(\vec{x}^{\prime}-\vec{x})
\end{equation}%
\begin{equation}
\left\{  \Phi_{a}(\vec{x}),\partial_{i}^{\prime}\Phi_{b}(\vec{x}^{\prime
})\right\}  =0\text{ \ };\text{ \ }\left\{  \Phi_{a}(\vec{x}),\Phi_{b}(\vec
{x}^{\prime})\right\}  =0.
\end{equation}
These are the first part of the well-known commutation relations of the Dirac
field. The others can be obtained using the adjoint Dirac equation
$i\partial_{\mu}\overline{\Psi}\gamma^{\mu}+m\overline{\Psi}=0.$

\section{ Conclusion}

\qquad The time covariance of the Dirac brackets has been successfully
demonstrated in the case of autonomous singular systems. This property made
possible the use of the CI method at the initial time, then directly
extrapolate the result to any later time. This way of proceeding is
straightforward, simple, and well suited to singular non-integrable systems.
The fundamental characteristic of this approach is that it does not require
the notion of constraints, nor their classification. This is a major
difference when compared to other alternatives.

Two cases, namely the self-dual model and the spinorial field, have been
successfully studied where the Dirac brackets were obtained directly from the
brackets of the initial conditions, which were calculated using a Taylor
expansion for the solution of the equations of motion.

This new approach is ready to be applied to any singular system to exploit its
simplicity and put its efficiency to the test.


\begin{thebibliography}{9}                                                                                                %


\bibitem {nous}Z. Belhadi, F. Menas, A. B\'{e}rard and H. Mohrbach, Annals of
Physics 351, 426--443 (2014).

\bibitem {Dirac}P.A.M. Dirac,\textit{\ Lectures on Quantum Mechanics}, Belfer
Graduate School of Science (1964).

\bibitem {erer}J. L. Anderson and P. G. Bergmann, Phys. Rev. \textbf{83, }1018 (1951).

\bibitem {FJ}L.\ Faddeev and R. Jackiw, Phys. Rev. Lett., 60\textbf{,} 1692 (1988).

\bibitem {town}P. K. Townsend, K. Pilch and P. Van Nieuwenhuizen, Phys. Let.,
136B, 38 (1984).

\bibitem {jk}S. Deser and R. Jackiw, Phys. Lett. B, 139(5-6), 371-373 (1984).

\bibitem {autre}H.O., Girotti, Int. J. Mod. Phys. A, 14(16), 2495-2510 (1999).

\bibitem {casa}R. Casalbuoni, Il Nuovo Cimento A, 33(1), 115-125 (1976).
\end{thebibliography}
\end{document}